\documentclass[aps,prb,twocolumn]{revtex4-2}
\usepackage{amsmath, amssymb}
\usepackage{graphicx}
\usepackage{xcolor}
\usepackage{ulem}
\usepackage[colorlinks,linkcolor=blue,anchorcolor=blue,citecolor=blue,urlcolor=blue]{hyperref}

\begin{document}

\title{Quantum interference among vortex bound states in superconductors}
\author{Yi-Ming Fu}
\author{Da Wang} \email{dawang@nju.edu.cn}
\author{Qiang-Hua Wang} \email{qhwang@nju.edu.cn}
\affiliation{National Laboratory of Solid State Microstructures $\&$ School of Physics, Nanjing University, Nanjing 210093, China}
\affiliation{Collaborative Innovation Center of Advanced Microstructures, Nanjing University, Nanjing 210093, China}

\begin{abstract}
In a recent experiment (Hou {\it et al.} Phys. Rev. X 15, 011027), a new type of necklace-like vortex bound state (VBS) was observed and attributed to disorder induced interference among different Caroli-de Gennes-Matricon (CdGM) states within one single vortex. In this work, we further investigate the possibilities of quantum interference among the CdGM states from different vortices in clean superconductors, which may become significant near the upper critical field. We find a series of interference patterns in the local density of states (LDOS) due to the overlap between spatially separated individual CdGM states. On a vortex lattice, the interference can also lead to a necklace-like LDOS, hence, providing an alternative and intrinsic mechanism to observe the novel necklace-like, or other spatially modulated VBS {  more generally}, in experiments. These results can be understood quite well within an effective tight-binding model constructed from the individual CdGM states, and can be checked in future experiments.
\end{abstract}

\maketitle

\section{Introduction}

In type-II superconductors, magnetic flux penetrates the material in the form of quantized vortices, each carrying a magnetic flux quantum $\Phi _0 = h/2e$ when the external magnetic field exceeds the lower critical field $H_{c1}$ and remains below the upper critical field $H_{c2}$.
Around a single vortex, the superconducting order parameter acquires a phase winding of $2\pi$ and drops to zero at the vortex center \cite{abrikosov_magnetic_1957}.
Such a topological defect acts as a potential well and hosts a series of discrete electronic vortex bound states (VBSs), called Caroli-de Gennes-Matricon (CdGM) states in a single vortex \cite{caroli_bound_1964}, with energy levels
$\sim\left( { l + \frac{1}{2}} \right)\Delta/k_F\xi$ where $\Delta$ is the superconducting gap far from the vortex, $k_F$ is the normal state Fermi momentum, $\xi$ is the superconducting coherence length, and $l$ is the angular momentum quantum number.
The CdGM wave function is $\left[J_l( k_F r)e^{-il\theta},J_{l+1}(k_F r)e^{-i(l+1)\theta} \right]$ times an radial envelope function which approaches $1$ for $r\ll\xi$ and decays exponentially for $r\gg\xi$.
The CdGM states have been confirmed in many scanning tunneling microscopy (STM) experiments \cite{hess_vortex-core_1990,guillamon_scanning_2008,song_direct_2011,hess_stm_1991,chen_discrete_2018,hess_scanning-tunneling-microscope_1989,rodrigo_scanning_2008,yin_scanning_2009,shan_observation_2011,zhou_visualizing_2013} by observing the zero-energy peak (due to finite resolution \cite{shore_density_1989}) or finite-energy ring-like local density of states (LDOS). In some cases, the CdGM states exhibit nematic or star-like feature due to the anisotropic Fermi surface or gap function \cite{Xiang_SCPMA_2024,nishimori_first_2004,chen_anisotropic_2023,liu_evidence_2019,chen_superconductivity_2023,hayashi_star-shaped_1996}.
In addition, the impurity effect on the CdGM states have also been investigated in some theoretical and experimental studies \cite{renner_scanning_1991,miranovic_effects_2004,masaki_impurity_2015,tsutsumi_coherence_2017,park_coherent_2021,de_mendonca_near_2023}.
{
On the other hand, in topological superconductors, the CdGM states exhibit a energy spectrum $\sim l\Delta/k_F\xi$. Notably, the zero-energy mode is a Majorana fermion which has attracted intensive research interest \cite{exp-FeTeSe,exp-LiFeOHFeSe,exp-LiFeAs} in recent years due to its potential applications in topological quantum computation \cite{Nayak2008}. }

Recently, Hou {\it et al.} \cite{hou_necklacelike_2025} reported the discovery of a new type of VBS exhibiting fast oscillatory ring shaped LDOS {on high-energy levels (approximately $0.6\Delta\sim 0.8\Delta$)}, dubbed as necklace-like VBS, in an iron based superconductor $\text{KCa}_2\text{Fe}_4\text{As}_4\text{F}_2$. This novel VBS was mainly attributed to disorder induced quantum interference between the CdGM states with opposite angular momenta $\pm l$ within one single vortex, leading to the necklace-like VBS by a second order perturbation.
But if there are two CdGM states with the same energy coming from adjacent vortices, they can interfere with each other in the leading order.
Actually, {on a vortex lattice,
the interference between neighboring vortices may cause observable effect in LDOS, which has not yet received much attention in the literature \cite{melnikov_electronic_2008,melnikov_local_2009,franz_quasiparticles_2000}.}

In this work, we are interested in studying the quantum interference between the CdGM states from different vortices. By solving the Bogoliubov-de Gennes (BdG) equations numerically on a lattice to obtain the VBSs in presence of two superconducting vortices or a triangular vortex lattice, we find a series of interference patterns in LDOS for the resulting VBSs. In particular, in an energy window with suitable inter-vortex distance, the interference can also lead to a necklace-like LDOS. { More generally,} this provides an additional mechanism to observe the modulated VBS in future experiments. Finally, we show that all these results can be understood quite well within an effective tight-binding model constructed from the isolated CdGM states as the Wannier basis.

\begin{figure*}
\includegraphics[width=17cm]{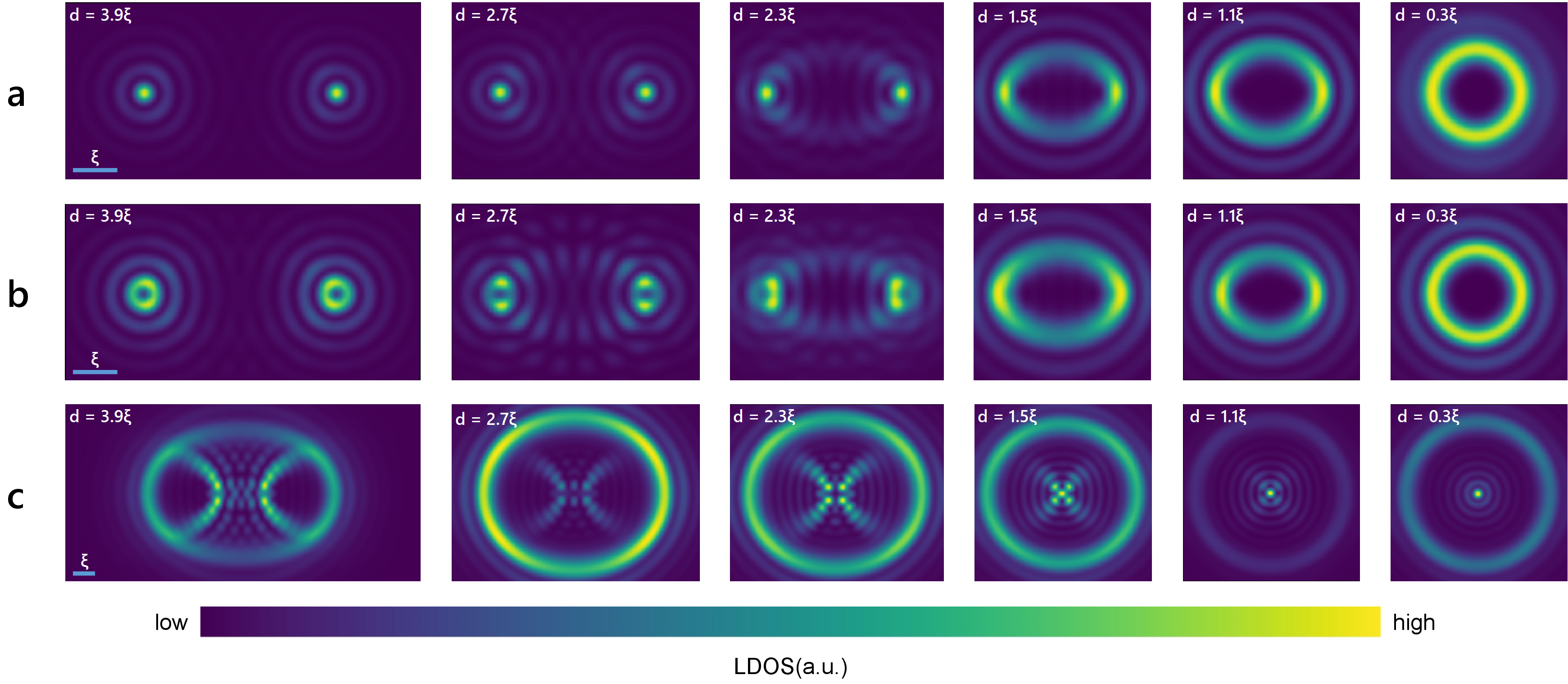}
\caption{LDOS mappings of two vortex cores gradually approaching each other at the $1$st (a), $3$rd (b) and $25$th (c) bound state energy levels. All the subfigures share the same scale bar (yellow for high and blue for low LDOS). The superconducting coherence length $\xi$ and the inter-vortex distance $d$ are labeled as indicated. {For the case of $d=0.3\xi$, the two vortices are equivalent to a double quanta vortex.} \label{fig1}}
\end{figure*}

\section{Model and Method}
To study the interference between different CdGM states, we adopt the following BdG Hamiltonian defined on a square lattice with spatially dependent onsite pairing potential,
\begin{align}
\hat H =&  - t\sum_{\left\langle {ij} \right\rangle \sigma } {\left( {{{\hat c}_{i\sigma }}^\dagger {{\hat c}_{j\sigma }} + {{\hat c}_{j\sigma }}^\dagger {{\hat c}_{i\sigma }}} \right)}  - {\mu}\sum\limits_{i\sigma } {{{\hat c}_{i\sigma }}^\dagger {{\hat c}_{i\sigma }}}  \nonumber\\
& + \sum_i {\left( {{\Delta _i}{{\hat c}_{i \uparrow }}^\dagger {{\hat c}_{i \downarrow }}^\dagger  + {\Delta _i}^* {{\hat c}_{i \downarrow }}{{\hat c}_{i \uparrow }}} \right)} , \label{eq1}
\end{align}
where $\hat{c}^{\dagger}_{i\sigma}$ creates an electron on site $i$ with spin $\sigma$, $t$ is the nearest neighbor hopping,
$\mu$ is the chemical potential to control the Fermi energy,
and $\Delta_i$ is the pairing order parameter.
For the case of a single vortex, we have
\begin{align}
{\Delta _i} = {\Delta _0}f\left( {{\vec r}_i - {\vec r}_{core}} \right),
\end{align}
where ${\vec r}_{core}$ is the coordinate of the vortex core and ${\vec r}_{i}$ is the coordinate of site $i$.
As usual, for simplicity, we choose $f\left( {\vec r} \right) = \tanh \left( r/\xi \right){e^{i\theta }}$
with $r$ the radial coordinate and $\theta$ the azimuthal angle of the position vector $\vec{r}$ \cite{Schopohl1995,hou_necklacelike_2025}.
For the case of two vortices, we choose the order parameter as \cite{melnikov_local_2009}
\begin{equation}
    {\Delta _i} = {\Delta _0}f\left( {{{\vec r}_i} - {{\vec r}_{core1}}} \right)f\left( {{{\vec r}_i} - {{\vec r}_{core2}}} \right) \label{eq2}
\end{equation}
where ${\vec r_{core1}},{\vec r_{core2}}$ are coordinates of the two vortex cores. Similarly, for the case of multiple vortex cores, we directly generalize it to
\begin{equation}
    {\Delta _i} = {\Delta _0}\prod\limits_{\vec{r}_{core}} {f\left( {{{\vec r}_i} - {{\vec r}_{core}}} \right)} \label{eq3}
\end{equation}
where the modulation comes from the product of all the vortex cores and thus enforces the phase singularities of $\Delta_i$ at all $\vec{r}_{core}$.

By exact diagonalization of the BdG Hamiltonian matrix in the basis of $[c_{i\uparrow},c_{i\downarrow}^\dag]^t$ on the $L\times L$ lattice, the bound state eigenvectors $[u_n,v_n]^t$ with eigen energies $|E_n|<\Delta_0$ can be obtained, from which the LDOS is given by
\begin{equation}
    \rho \left( {i,\omega } \right) = \frac{2}{\pi }\sum\limits_n {\frac{{{{\left| {{u_n}\left( i \right)} \right|}^2}\eta }}{{{{\left( {\omega  - {E_n}} \right)}^2} + {\eta ^2}}}} \label{eq4}
\end{equation}
where $\eta$ is the smearing factor corresponding to the energy level expansion. In calculations below, unless specified otherwise, we set
${\mu _0} =  - 3.5t$, ${\Delta _0} = 0.1t$, $\xi  = 10$, $L = 200$, $\eta=0.001t$. With these parameters, we have the dimensionless quantity $k_F\xi\sim7$, which means that we are far from the quasi-classical limit such that the discrete VBSs can be identified.

\begin{figure*}
\includegraphics[width=\linewidth]{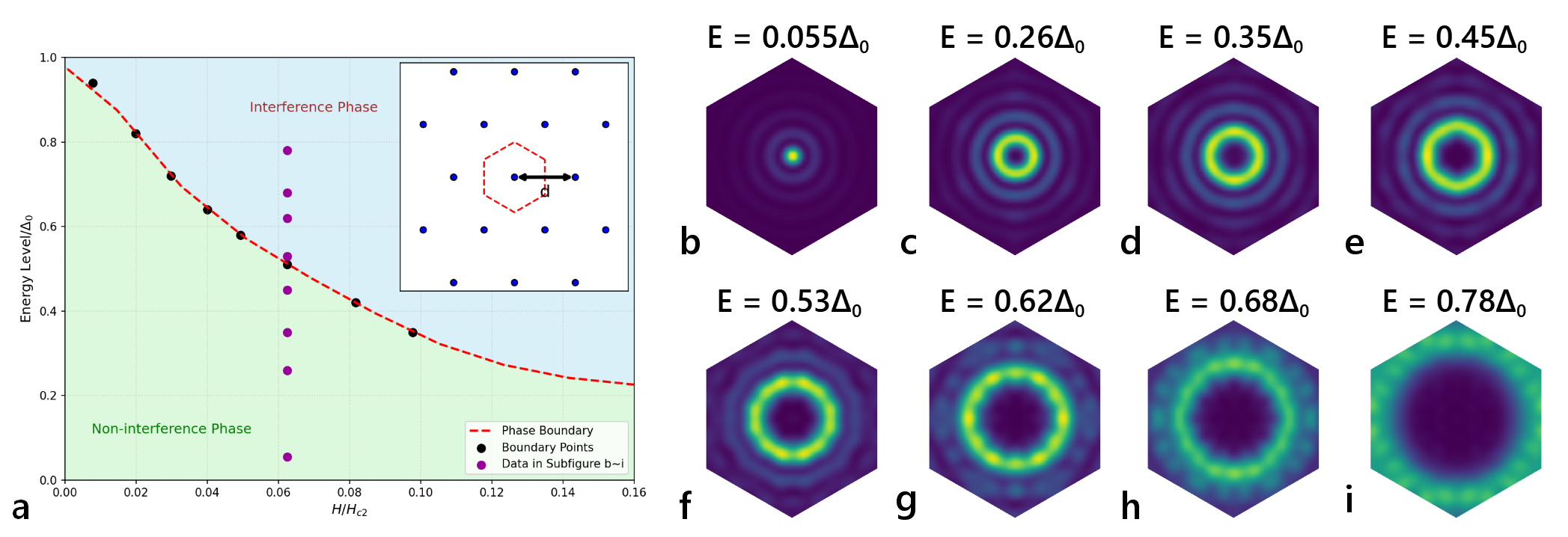}
\caption{(a) {presents a phase diagram on the plane of energy and magnetic field, indicating where the LDOS maps exhibit quantum interference patterns. A schematic diagram of the hexagonal vortex lattice is given in the inset.} In the Wigner-Seitz cell as enclosed by the red dashed regular hexagon, the LDOS for different bound state energies {and inter-vortex distance $d=4\xi$} are displayed from (b) to (i). All the LDOS maps share the same scale bar similar to Fig.~\ref{fig1}. \label{fig2}
}
\end{figure*}

\section{Results and Discussions}

\subsection{vortex pair}
Let us first look at the results of a pair of vortices. In Fig.~\ref{fig1}, we present the LDOS mappings around these two vortices as the distance $d$ between their centers gradually reduces (as indicated), at the first VBS level $E_1$ ($0\sim0.056\Delta_0$) in (a), the third level $E_3$ ($0.12\sim 0.16{\Delta _0}$) in (b), and the 25th level $E_{25}$ ($0.70\sim 0.76{\Delta _0}$) in (c), respectively.
For large $d\gg\xi$, the CdGM states of the two vortices are spatially well separated and thus do not interact with each other, as shown in Fig.~\ref{fig1}(a) and (b) for $d=3.9\xi$.
As $d$ gradually decreases, the CdGM state wave functions of the two vortices overlap with each other and generate a series of interference patterns. Interestingly, as $d$ further decreases to be much smaller $\xi$, say $0.3\xi$ as displayed, the resulting VBS becomes almost ring-like, even for the bound state with the lowest energy.
This is quite different from the single quantum vortex (point-like), but agrees well with the so-called multi-quantum vortex or giant vortex \cite{melnikov_local_2009,silaev_self-consistent_2013}.

Another way to enhance the interference is to increase the spatial size of the individual CdGM wave functions, corresponding to higher energy states. This is actually seen in Fig.~\ref{fig1}(c) for the 25th bound states. For $d=3.9\xi$, the low energy levels ($E_1$ and $E_3$) exhibit little interference effect but the $25$th level $E_{25}$ displays a significant interference pattern. In particular, in the middle region between the two vortex centers, the oscillated LDOS is clearly seen. Instead, in the outer region of the two vortices, only an elliptic ring is seen but without fast oscillation. These features can be understood as follows. In the middle region, the phases of the two individual CdGM wave functions increase towards opposite directions, resulting in fast oscillation by their interference. But in the outer region, the phases of the two CdGM wave functions are almost synchronized, only resulting in a interference with long wave length like the beat phenomenon.
As the inter-vortex distance $d$ decreases, the interference pattern in the middle region shrinks and vanishes finally for $d\ll\xi$. Meanwhile, the outer elliptic ring becomes more circular and more uniform on the ring, similar to $E_1$ and $E_3$, as a manifestation of the vortex bound states inside a giant vortex.

\begin{figure*}
\includegraphics[width=\linewidth]{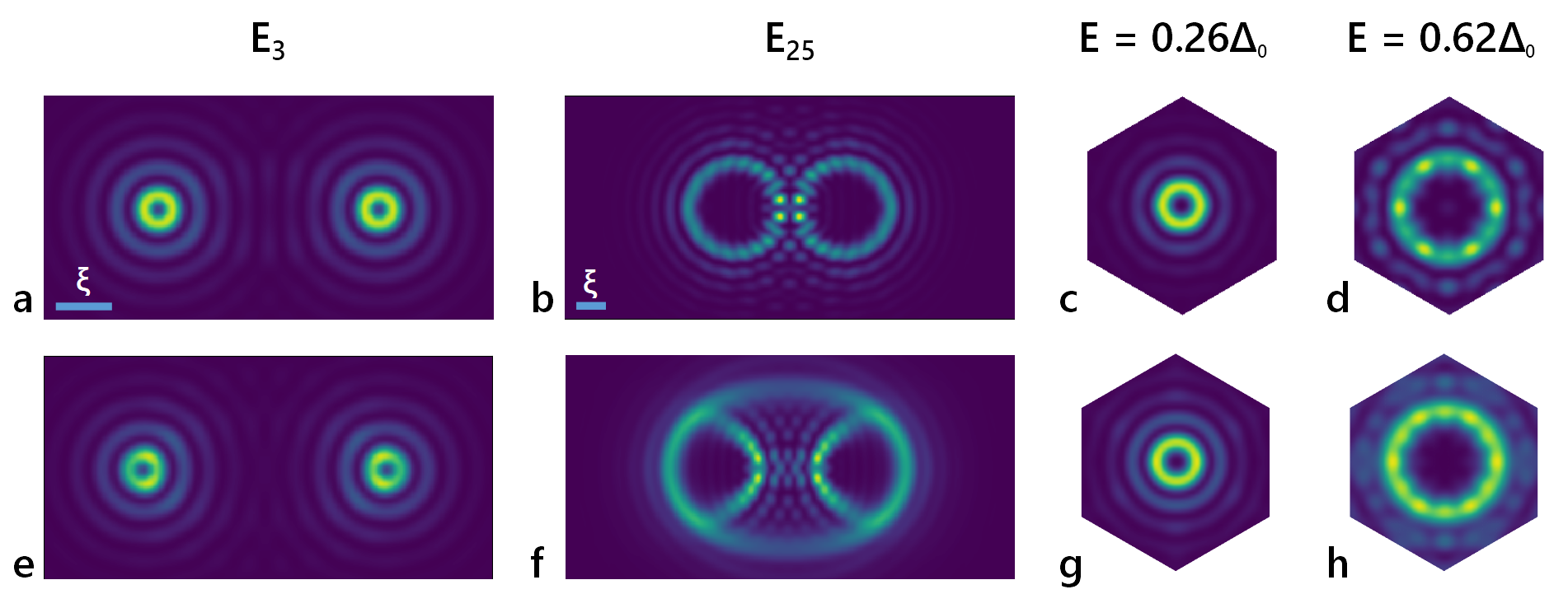}
\caption{LDOS by the tight-binding model for a vortex pair with $d=3.9\xi$ at $E_3$ (a) and $E_{25}$ (b), for a vortex lattice at $0.26\Delta_0$ (c) and $0.62\Delta_0$ (d).
As comparisons, the corresponding results by exact diagonalization are given in (e-h). The color bar is similar to Fig.~\ref{fig1}. \label{fig3}}
\end{figure*}

\subsection{vortex lattice}
From the above results, we have learned that the high energy CdGM states can interfere with each other to produce an interference pattern in the middle region between these two vortices.
This motivates us to further investigate the vortex lattice to see whether the above interference segment can be connected to become complete, since there are more than one vortices around a central one.
In the following, we consider the hexagonal vortex lattice as shown in {the inset of} Fig.~\ref{fig2}(a) with the vortex lattice constant $d=4\xi$.
The LDOS around one vortex for a series of energies are presented in Fig.~\ref{fig2}(b-i).
As anticipated, the low energy VBSs, e.g. Fig.~\ref{fig2}(b-d), are similar to the individual CdGM states of a single vortex and the vortex lattice has little impact on these states because they are more local and thus have little overlap with each other.
But as the bound state energy increases, the size of the CdGM states increases. For the present value of inter-vortex distance $d=4\xi$, we find as $E$ approaches $0.5\Delta_0$, the spatial size of the CdGM states are comparable to $d/2=2\xi$ such that they begin to interfere with each other. The resulting VBS exhibits an obvious sixfold rotational symmetry as in Fig.~\ref{fig2}(e) and then displays the complete necklace-like oscillation as shown in Fig.~\ref{fig2}(f-i). Note that the oscillation pattern becomes clearest for a particular energy, here with $E=0.62\Delta_0$ as shown in Fig.~\ref{fig2}(g), when the neighboring CdGM wave functions overlap significantly.
Further increasing the bound state energy suppresses the necklace-like oscillation as shown in Fig.~\ref{fig2}(h) and (i).
We have also studied the square vortex lattice and found that the results (not shown) are similar to those in the triangular lattice.
{By varying the inter-vortex distance $d\approx\xi/\sqrt{H/H_{c2}}$, we map out the phase diagram versus $H/H_{c2}$ in Fig.~\ref{fig2}(a). As increasing the magnetic field $H$, the inter-vortex distance is reduced, such that the quantum interference occurs for lower-energy CdGM states.
From this phase diagram, we expect the zero-energy CdGM state (not here but in p-wave superconductors), as a Majorana zero mode, is stable against the quantum interference, since the Majorana wave function is always real up to a global phase.
}

Therefore, the quantum interference of adjacent CdGM states on a vortex lattice can yield the necklace-like VBS under particular conditions with suitable inter-vortex distance and bound state energies, {  or various interference patterns more generally.
In particular, in this mechanism, quantum interference occurs for high energy VBS, which is also consistent with Ref.~\onlinecite{hou_necklacelike_2025}.
Therefore,} this provides a new way, as an intrinsic mechanism without the need of disorder, to observe the necklace-like or other spatially modulated VBS in future experiments.
the intrinsic inter-vortex interference here and the extrinsic disorder mechanism can be distinguished easily: the oscillation phase is bound to the vortex lattice direction in the intrinsic mechanism, but not in the disorder mechanism.

%As shown in Figure \ref{fig2}b$\sim$i, within this vortex lattice configuration, the coherence between VBS states on different vortex cores is not significant at lower energy levels, and the periodic arrangement of vortex cores has minimal impact on the LDOS patterns of VBS. However, when the energy reaches approximately $0.5\Delta_0$
%, the intrinsic diameter of the VBS associated with a single vortex core is close to the distance between two nearest-neighbor vortex cores, which make the coherence between VBS states on different vortex cores become significant, and this periodic and symmetric arrangement induces a distinct necklace-like pattern on the whole LDOS ring, especially for
%$E=0.62 \Delta_0$.
%For higher-energy CdGM states, if their characteristic spatial extent exceeds the lattice spacing of the vortex core lattice, individual CdGM states will not be resolvable at the lattice center, as demonstrated in Figure \ref{fig2}i.

\subsection{effective tight-binding model}

We have seen the quantum interference among the CdGM states of different vortices can lead to a series of interference patterns.
Now we construct an effective tight-binding model to describe such an interference more quantitatively.
We first choose the non-interacting individual CdGM states $u_n(\vec{r}-\vec{r}_i)$ (by solving the single vortex) as the Wannier functions by assuming their orthogonality for sufficiently large inter-vortex distance. Second, because the individual CdGM states have different energies $E_n$ with adjacent gaps of order $\sim\Delta_0/k_F\xi$, we simply assume the hoppings between these states on the neighboring sites (vortex centers) are dominated by those with the same energy $E_n$. That is, the effective model is block diagonalized in $n$-space, given by
\begin{equation}
    {\hat H_{\rm eff} } = \sum\limits_{ni} {{E_n}{{\hat \alpha}_{ni}}^\dagger {{\hat \alpha}_{ni}}}  + \sum\limits_{n\left\langle {ij} \right\rangle } {{t_n}\left( {{{\hat \alpha}_{ni}}^\dagger {{\hat \alpha}_{nj}} + {{\hat \alpha}_{nj}}^\dagger {{\hat \alpha}_{ni}}} \right)} , \label{eq5}
\end{equation}
where $\hat{\alpha}_{ni}^\dag$ creates a CdGM state with energy $E_n$ at site (vortex center) $i$, and $t_n$ is the nearest neighbor hopping between two CdGM states. Due to the chiral property of the CdGM states, the hopping $t_n$ along different directional bonds are the same. Of course, there can be a phase freedom in $t_n$, which brings a phase shift to the hybridized CdGM states. But we numerically find the phase of $t_n$ is always zero to match the result of exact diagonalization.

By solving the tight-binding model, one can obtain the $k$-th eigenstate $a_{nk}(i)$. Then the resulting VBS wave function $\tilde{u}_{nk}$ is given by $\tilde{u}_{nk}(\vec{r})=\sum_i a_{nk}(i)u_{n}(\vec{r}-\vec{r}_i)$.
In Fig.~\ref{fig3}, we compare the results constructed by the effective tight-binding model with the results obtained by exact diagonalization in the above.
For the case of a vortex pair, the tight-binding model simply implies bonding and antibonding of the two CdGM states as the hybridized VBS states. In Fig.~\ref{fig3}(a) and (b), we present the tight-binding wave functions for the $3$rd and $25$-th energy levels. The corresponding results obtained by exact diagonalization are also given in Fig.~\ref{fig3}(e) and (f), respectively. As can be seen, the main interference patterns particularly in the middle region of the two vortices are similar to the exact results, although the outer ring for $E_{25}$ is not captured by the tight-binding approximation.

Applying the tight-binding method to the vortex lattice, since the dominated interference occurs in the middle region between each two neighboring vortices, the tight-binding approximation is expected to  work better than the case of the vortex pair. For the vortex lattice, the results of LDOS at $E=0.26\Delta_0$ and $0.62\Delta_0$ are shown in Fig.~\ref{fig3}(c) and (d). Clearly, they are indeed in good agreement with the exact results as given in Fig.~\ref{fig3}(g) and (h), respectively.
Such a consistence confirms our understanding of the new and intrinsic mechanism for the necklace-like or other spatially modulated VBS caused by quantum interference among the CdGM states of different vortices.

\section{Summary}
In summary, through exact diagonalization of the BdG Hamiltonian, we numerically investigated the evolution of LDOS patterns versus energy for vortex bound states for both two vortices and vortex lattice.
We find that even in clean superconductors, quantum interference between individual CdGM states localized at different vortex cores can also lead to the necklace-like or other spatially modulated vortex bound state patterns,
which is a supplement to the disorder mechanism to further investigate the quantum interference phenomena in vortex bound states.

\begin{acknowledgments}
This work is supported by National Key R\&D Program of China (Grant No. 2022YFA1403201), National Natural Science Foundation of China (Grant No. 12274205, No. 12374147 and No. 92365203).
\end{acknowledgments}

\bibliography{00.bib}

\begin{thebibliography}{35}%
\makeatletter
\providecommand \@ifxundefined [1]{%
 \@ifx{#1\undefined}
}%
\providecommand \@ifnum [1]{%
 \ifnum #1\expandafter \@firstoftwo
 \else \expandafter \@secondoftwo
 \fi
}%
\providecommand \@ifx [1]{%
 \ifx #1\expandafter \@firstoftwo
 \else \expandafter \@secondoftwo
 \fi
}%
\providecommand \natexlab [1]{#1}%
\providecommand \enquote  [1]{``#1''}%
\providecommand \bibnamefont  [1]{#1}%
\providecommand \bibfnamefont [1]{#1}%
\providecommand \citenamefont [1]{#1}%
\providecommand \href@noop [0]{\@secondoftwo}%
\providecommand \href [0]{\begingroup \@sanitize@url \@href}%
\providecommand \@href[1]{\@@startlink{#1}\@@href}%
\providecommand \@@href[1]{\endgroup#1\@@endlink}%
\providecommand \@sanitize@url [0]{\catcode `\\12\catcode `\$12\catcode
  `\&12\catcode `\#12\catcode `\^12\catcode `\_12\catcode `\%12\relax}%
\providecommand \@@startlink[1]{}%
\providecommand \@@endlink[0]{}%
\providecommand \url  [0]{\begingroup\@sanitize@url \@url }%
\providecommand \@url [1]{\endgroup\@href {#1}{\urlprefix }}%
\providecommand \urlprefix  [0]{URL }%
\providecommand \Eprint [0]{\href }%
\providecommand \doibase [0]{https://doi.org/}%
\providecommand \selectlanguage [0]{\@gobble}%
\providecommand \bibinfo  [0]{\@secondoftwo}%
\providecommand \bibfield  [0]{\@secondoftwo}%
\providecommand \translation [1]{[#1]}%
\providecommand \BibitemOpen [0]{}%
\providecommand \bibitemStop [0]{}%
\providecommand \bibitemNoStop [0]{.\EOS\space}%
\providecommand \EOS [0]{\spacefactor3000\relax}%
\providecommand \BibitemShut  [1]{\csname bibitem#1\endcsname}%
\let\auto@bib@innerbib\@empty
%</preamble>
\bibitem [{\citenamefont {Hou}\ \emph {et~al.}(2025)\citenamefont {Hou},
  \citenamefont {Chen}, \citenamefont {Hong}, \citenamefont {Wang},
  \citenamefont {Duan}, \citenamefont {Yang}, \citenamefont {Li}, \citenamefont
  {Luo}, \citenamefont {Wang}, \citenamefont {Xiang},\ and\ \citenamefont
  {Wen}}]{hou_necklacelike_2025}%
  \BibitemOpen
  \bibfield  {author} {\bibinfo {author} {\bibfnamefont {Z.}~\bibnamefont
  {Hou}}, \bibinfo {author} {\bibfnamefont {K.}~\bibnamefont {Chen}}, \bibinfo
  {author} {\bibfnamefont {W.}~\bibnamefont {Hong}}, \bibinfo {author}
  {\bibfnamefont {D.}~\bibnamefont {Wang}}, \bibinfo {author} {\bibfnamefont
  {W.}~\bibnamefont {Duan}}, \bibinfo {author} {\bibfnamefont {H.}~\bibnamefont
  {Yang}}, \bibinfo {author} {\bibfnamefont {S.}~\bibnamefont {Li}}, \bibinfo
  {author} {\bibfnamefont {H.}~\bibnamefont {Luo}}, \bibinfo {author}
  {\bibfnamefont {Q.-H.}\ \bibnamefont {Wang}}, \bibinfo {author}
  {\bibfnamefont {T.}~\bibnamefont {Xiang}},\ and\ \bibinfo {author}
  {\bibfnamefont {H.-H.}\ \bibnamefont {Wen}},\ }\bibfield  {title} {\bibinfo
  {title} {Necklacelike {Pattern} of {Vortex} {Bound} {States}},\ }\href
  {https://doi.org/10.1103/PhysRevX.15.011027} {\bibfield  {journal} {\bibinfo
  {journal} {Phys. Rev. X}\ }\textbf {\bibinfo {volume} {15}},\ \bibinfo
  {pages} {011027} (\bibinfo {year} {2025})}\BibitemShut {NoStop}%
\bibitem [{\citenamefont {Abrikosov}(1957)}]{abrikosov_magnetic_1957}%
  \BibitemOpen
  \bibfield  {author} {\bibinfo {author} {\bibfnamefont {A.~A.}\ \bibnamefont
  {Abrikosov}},\ }\bibfield  {title} {\bibinfo {title} {On the magnetic
  properties of superconductors of the second group},\ }\href
  {https://webofscience.clarivate.cn/wos/woscc/full-record/WOS:A1957WN04300022}
  {\bibfield  {journal} {\bibinfo  {journal} {Sov. Phys. JETP}\ }\textbf
  {\bibinfo {volume} {5}},\ \bibinfo {pages} {1174} (\bibinfo {year}
  {1957})}\BibitemShut {NoStop}%
\bibitem [{\citenamefont {Caroli}\ \emph {et~al.}(1964)\citenamefont {Caroli},
  \citenamefont {De~Gennes},\ and\ \citenamefont
  {Matricon}}]{caroli_bound_1964}%
  \BibitemOpen
  \bibfield  {author} {\bibinfo {author} {\bibfnamefont {C.}~\bibnamefont
  {Caroli}}, \bibinfo {author} {\bibfnamefont {P.~G.}\ \bibnamefont
  {De~Gennes}},\ and\ \bibinfo {author} {\bibfnamefont {J.}~\bibnamefont
  {Matricon}},\ }\bibfield  {title} {\bibinfo {title} {Bound {Fermion} states
  on a vortex line in a type {II} superconductor},\ }\href
  {https://doi.org/10.1016/0031-9163(64)90375-0} {\bibfield  {journal}
  {\bibinfo  {journal} {Phys. Lett.}\ }\textbf {\bibinfo {volume} {9}},\
  \bibinfo {pages} {307} (\bibinfo {year} {1964})}\BibitemShut {NoStop}%
\bibitem [{\citenamefont {Hess}\ \emph {et~al.}(1990)\citenamefont {Hess},
  \citenamefont {Robinson},\ and\ \citenamefont
  {Waszczak}}]{hess_vortex-core_1990}%
  \BibitemOpen
  \bibfield  {author} {\bibinfo {author} {\bibfnamefont {H.~F.}\ \bibnamefont
  {Hess}}, \bibinfo {author} {\bibfnamefont {R.~B.}\ \bibnamefont {Robinson}},\
  and\ \bibinfo {author} {\bibfnamefont {J.~V.}\ \bibnamefont {Waszczak}},\
  }\bibfield  {title} {\bibinfo {title} {Vortex-core structure observed with a
  scanning tunneling microscope},\ }\href
  {https://doi.org/10.1103/PhysRevLett.64.2711} {\bibfield  {journal} {\bibinfo
   {journal} {Phys. Rev. Lett.}\ }\textbf {\bibinfo {volume} {64}},\ \bibinfo
  {pages} {2711} (\bibinfo {year} {1990})}\BibitemShut {NoStop}%
\bibitem [{\citenamefont {Guillamon}\ \emph {et~al.}(2008)\citenamefont
  {Guillamon}, \citenamefont {Suderow}, \citenamefont {Vieira},\ and\
  \citenamefont {Rodiere}}]{guillamon_scanning_2008}%
  \BibitemOpen
  \bibfield  {author} {\bibinfo {author} {\bibfnamefont {I.}~\bibnamefont
  {Guillamon}}, \bibinfo {author} {\bibfnamefont {H.}~\bibnamefont {Suderow}},
  \bibinfo {author} {\bibfnamefont {S.}~\bibnamefont {Vieira}},\ and\ \bibinfo
  {author} {\bibfnamefont {P.}~\bibnamefont {Rodiere}},\ }\bibfield  {title}
  {\bibinfo {title} {Scanning tunneling spectroscopy with superconducting tips
  of {Al}},\ }\href {https://doi.org/10.1016/j.physc.2007.11.066} {\bibfield
  {journal} {\bibinfo  {journal} {Phys. C: Supercond. Appl.}\ }\textbf
  {\bibinfo {volume} {468}},\ \bibinfo {pages} {537} (\bibinfo {year}
  {2008})}\BibitemShut {NoStop}%
\bibitem [{\citenamefont {Song}\ \emph {et~al.}(2011)\citenamefont {Song},
  \citenamefont {Wang}, \citenamefont {Cheng}, \citenamefont {Jiang},
  \citenamefont {Li}, \citenamefont {Zhang}, \citenamefont {Li}, \citenamefont
  {He}, \citenamefont {Wang}, \citenamefont {Jia}, \citenamefont {Hung},
  \citenamefont {Wu}, \citenamefont {Ma}, \citenamefont {Chen},\ and\
  \citenamefont {Xue}}]{song_direct_2011}%
  \BibitemOpen
  \bibfield  {author} {\bibinfo {author} {\bibfnamefont {C.-L.}\ \bibnamefont
  {Song}}, \bibinfo {author} {\bibfnamefont {Y.-L.}\ \bibnamefont {Wang}},
  \bibinfo {author} {\bibfnamefont {P.}~\bibnamefont {Cheng}}, \bibinfo
  {author} {\bibfnamefont {Y.-P.}\ \bibnamefont {Jiang}}, \bibinfo {author}
  {\bibfnamefont {W.}~\bibnamefont {Li}}, \bibinfo {author} {\bibfnamefont
  {T.}~\bibnamefont {Zhang}}, \bibinfo {author} {\bibfnamefont
  {Z.}~\bibnamefont {Li}}, \bibinfo {author} {\bibfnamefont {K.}~\bibnamefont
  {He}}, \bibinfo {author} {\bibfnamefont {L.}~\bibnamefont {Wang}}, \bibinfo
  {author} {\bibfnamefont {J.-F.}\ \bibnamefont {Jia}}, \bibinfo {author}
  {\bibfnamefont {H.-H.}\ \bibnamefont {Hung}}, \bibinfo {author}
  {\bibfnamefont {C.}~\bibnamefont {Wu}}, \bibinfo {author} {\bibfnamefont
  {X.}~\bibnamefont {Ma}}, \bibinfo {author} {\bibfnamefont {X.}~\bibnamefont
  {Chen}},\ and\ \bibinfo {author} {\bibfnamefont {Q.-K.}\ \bibnamefont
  {Xue}},\ }\bibfield  {title} {\bibinfo {title} {Direct {Observation} of
  {Nodes} and {Twofold} {Symmetry} in {FeSe} {Superconductor}},\ }\href
  {https://doi.org/10.1126/science.1202226} {\bibfield  {journal} {\bibinfo
  {journal} {Science}\ }\textbf {\bibinfo {volume} {332}},\ \bibinfo {pages}
  {1410} (\bibinfo {year} {2011})}\BibitemShut {NoStop}%
\bibitem [{\citenamefont {Hess}\ \emph {et~al.}(1991)\citenamefont {Hess},
  \citenamefont {Robinson},\ and\ \citenamefont {Waszczak}}]{hess_stm_1991}%
  \BibitemOpen
  \bibfield  {author} {\bibinfo {author} {\bibfnamefont {H.~F.}\ \bibnamefont
  {Hess}}, \bibinfo {author} {\bibfnamefont {R.~B.}\ \bibnamefont {Robinson}},\
  and\ \bibinfo {author} {\bibfnamefont {J.~V.}\ \bibnamefont {Waszczak}},\
  }\bibfield  {title} {\bibinfo {title} {{STM} spectroscopy of vortex cores and
  the flux lattice},\ }\href {https://doi.org/10.1016/0921-4526(91)90262-D}
  {\bibfield  {journal} {\bibinfo  {journal} {Physica B: Condensed Matter}\
  }\textbf {\bibinfo {volume} {169}},\ \bibinfo {pages} {422} (\bibinfo {year}
  {1991})}\BibitemShut {NoStop}%
\bibitem [{\citenamefont {Chen}\ \emph {et~al.}(2018)\citenamefont {Chen},
  \citenamefont {Chen}, \citenamefont {Yang}, \citenamefont {Du}, \citenamefont
  {Zhu}, \citenamefont {Wang},\ and\ \citenamefont {Wen}}]{chen_discrete_2018}%
  \BibitemOpen
  \bibfield  {author} {\bibinfo {author} {\bibfnamefont {M.}~\bibnamefont
  {Chen}}, \bibinfo {author} {\bibfnamefont {X.}~\bibnamefont {Chen}}, \bibinfo
  {author} {\bibfnamefont {H.}~\bibnamefont {Yang}}, \bibinfo {author}
  {\bibfnamefont {Z.}~\bibnamefont {Du}}, \bibinfo {author} {\bibfnamefont
  {X.}~\bibnamefont {Zhu}}, \bibinfo {author} {\bibfnamefont {E.}~\bibnamefont
  {Wang}},\ and\ \bibinfo {author} {\bibfnamefont {H.-H.}\ \bibnamefont
  {Wen}},\ }\bibfield  {title} {\bibinfo {title} {Discrete energy levels of
  {Caroli}-de {Gennes}-{Matricon} states in quantum limit in
  fete$_{0.55}$se$_{0.45}$},\ }\href
  {https://doi.org/10.1038/s41467-018-03404-8} {\bibfield  {journal} {\bibinfo
  {journal} {Nature Communications}\ }\textbf {\bibinfo {volume} {9}},\
  \bibinfo {pages} {970} (\bibinfo {year} {2018})}\BibitemShut {NoStop}%
\bibitem [{\citenamefont {Hess}\ \emph {et~al.}(1989)\citenamefont {Hess},
  \citenamefont {Robinson}, \citenamefont {Dynes}, \citenamefont {Valles},\
  and\ \citenamefont {Waszczak}}]{hess_scanning-tunneling-microscope_1989}%
  \BibitemOpen
  \bibfield  {author} {\bibinfo {author} {\bibfnamefont {H.~F.}\ \bibnamefont
  {Hess}}, \bibinfo {author} {\bibfnamefont {R.~B.}\ \bibnamefont {Robinson}},
  \bibinfo {author} {\bibfnamefont {R.~C.}\ \bibnamefont {Dynes}}, \bibinfo
  {author} {\bibfnamefont {J.~M.}\ \bibnamefont {Valles}},\ and\ \bibinfo
  {author} {\bibfnamefont {J.~V.}\ \bibnamefont {Waszczak}},\ }\bibfield
  {title} {\bibinfo {title} {Scanning-{Tunneling}-{Microscope} {Observation} of
  the {Abrikosov} {Flux} {Lattice} and the {Density} of {States} near and
  inside a {Fluxoid}},\ }\href {https://doi.org/10.1103/PhysRevLett.62.214}
  {\bibfield  {journal} {\bibinfo  {journal} {Physical Review Letters}\
  }\textbf {\bibinfo {volume} {62}},\ \bibinfo {pages} {214} (\bibinfo {year}
  {1989})}\BibitemShut {NoStop}%
\bibitem [{\citenamefont {Rodrigo}\ \emph {et~al.}(2008)\citenamefont
  {Rodrigo}, \citenamefont {Crespo},\ and\ \citenamefont
  {Vieira}}]{rodrigo_scanning_2008}%
  \BibitemOpen
  \bibfield  {author} {\bibinfo {author} {\bibfnamefont {J.~G.}\ \bibnamefont
  {Rodrigo}}, \bibinfo {author} {\bibfnamefont {V.}~\bibnamefont {Crespo}},\
  and\ \bibinfo {author} {\bibfnamefont {S.}~\bibnamefont {Vieira}},\
  }\bibfield  {title} {\bibinfo {title} {Scanning tunneling spectroscopy of the
  vortex state in {NbSe2} using a superconducting tip},\ }\href
  {https://doi.org/10.1016/j.physc.2007.11.019} {\bibfield  {journal} {\bibinfo
   {journal} {Physica C: Superconductivity and its Applications}\ }\bibinfo
  {series} {Proceedings of the {Fifth} {International} {Conference} on {Vortex}
  {Matter} in {Nanostructured} {Superconductors}},\ \textbf {\bibinfo {volume}
  {468}},\ \bibinfo {pages} {547} (\bibinfo {year} {2008})}\BibitemShut
  {NoStop}%
\bibitem [{\citenamefont {Yin}\ \emph {et~al.}(2009)\citenamefont {Yin},
  \citenamefont {Zech}, \citenamefont {Williams}, \citenamefont {Wang},
  \citenamefont {Wu}, \citenamefont {Chen},\ and\ \citenamefont
  {Hoffman}}]{yin_scanning_2009}%
  \BibitemOpen
  \bibfield  {author} {\bibinfo {author} {\bibfnamefont {Y.}~\bibnamefont
  {Yin}}, \bibinfo {author} {\bibfnamefont {M.}~\bibnamefont {Zech}}, \bibinfo
  {author} {\bibfnamefont {T.~L.}\ \bibnamefont {Williams}}, \bibinfo {author}
  {\bibfnamefont {X.~F.}\ \bibnamefont {Wang}}, \bibinfo {author}
  {\bibfnamefont {G.}~\bibnamefont {Wu}}, \bibinfo {author} {\bibfnamefont
  {X.~H.}\ \bibnamefont {Chen}},\ and\ \bibinfo {author} {\bibfnamefont
  {J.~E.}\ \bibnamefont {Hoffman}},\ }\bibfield  {title} {\bibinfo {title}
  {Scanning {Tunneling} {Spectroscopy} and {Vortex} {Imaging} in the {Iron}
  {Pnictide} {Superconductor} bafe$_{1.8}$co$_{0.2}$as$_2$},\ }\href
  {https://doi.org/10.1103/PhysRevLett.102.097002} {\bibfield  {journal}
  {\bibinfo  {journal} {Physical Review Letters}\ }\textbf {\bibinfo {volume}
  {102}},\ \bibinfo {pages} {097002} (\bibinfo {year} {2009})}\BibitemShut
  {NoStop}%
\bibitem [{\citenamefont {Shan}\ \emph {et~al.}(2011)\citenamefont {Shan},
  \citenamefont {Wang}, \citenamefont {Shen}, \citenamefont {Zeng},
  \citenamefont {Huang}, \citenamefont {Li}, \citenamefont {Wang},
  \citenamefont {Yang}, \citenamefont {Ren}, \citenamefont {Wang},
  \citenamefont {Pan},\ and\ \citenamefont {Wen}}]{shan_observation_2011}%
  \BibitemOpen
  \bibfield  {author} {\bibinfo {author} {\bibfnamefont {L.}~\bibnamefont
  {Shan}}, \bibinfo {author} {\bibfnamefont {Y.-L.}\ \bibnamefont {Wang}},
  \bibinfo {author} {\bibfnamefont {B.}~\bibnamefont {Shen}}, \bibinfo {author}
  {\bibfnamefont {B.}~\bibnamefont {Zeng}}, \bibinfo {author} {\bibfnamefont
  {Y.}~\bibnamefont {Huang}}, \bibinfo {author} {\bibfnamefont
  {A.}~\bibnamefont {Li}}, \bibinfo {author} {\bibfnamefont {D.}~\bibnamefont
  {Wang}}, \bibinfo {author} {\bibfnamefont {H.}~\bibnamefont {Yang}}, \bibinfo
  {author} {\bibfnamefont {C.}~\bibnamefont {Ren}}, \bibinfo {author}
  {\bibfnamefont {Q.-H.}\ \bibnamefont {Wang}}, \bibinfo {author}
  {\bibfnamefont {S.~H.}\ \bibnamefont {Pan}},\ and\ \bibinfo {author}
  {\bibfnamefont {H.-H.}\ \bibnamefont {Wen}},\ }\bibfield  {title} {\bibinfo
  {title} {Observation of ordered vortices with {Andreev} bound states in
  ba$_{0.6}$k$_{0.4}$fe$_2$as$_2$},\ }\href {https://doi.org/10.1038/nphys1908}
  {\bibfield  {journal} {\bibinfo  {journal} {Nature Physics}\ }\textbf
  {\bibinfo {volume} {7}},\ \bibinfo {pages} {325} (\bibinfo {year}
  {2011})}\BibitemShut {NoStop}%
\bibitem [{\citenamefont {Zhou}\ \emph {et~al.}(2013)\citenamefont {Zhou},
  \citenamefont {Misra}, \citenamefont {da~Silva~Neto}, \citenamefont
  {Aynajian}, \citenamefont {Baumbach}, \citenamefont {Thompson}, \citenamefont
  {Bauer},\ and\ \citenamefont {Yazdani}}]{zhou_visualizing_2013}%
  \BibitemOpen
  \bibfield  {author} {\bibinfo {author} {\bibfnamefont {B.~B.}\ \bibnamefont
  {Zhou}}, \bibinfo {author} {\bibfnamefont {S.}~\bibnamefont {Misra}},
  \bibinfo {author} {\bibfnamefont {E.~H.}\ \bibnamefont {da~Silva~Neto}},
  \bibinfo {author} {\bibfnamefont {P.}~\bibnamefont {Aynajian}}, \bibinfo
  {author} {\bibfnamefont {R.~E.}\ \bibnamefont {Baumbach}}, \bibinfo {author}
  {\bibfnamefont {J.~D.}\ \bibnamefont {Thompson}}, \bibinfo {author}
  {\bibfnamefont {E.~D.}\ \bibnamefont {Bauer}},\ and\ \bibinfo {author}
  {\bibfnamefont {A.}~\bibnamefont {Yazdani}},\ }\bibfield  {title} {\bibinfo
  {title} {Visualizing nodal heavy fermion superconductivity in cecoin$_5$},\
  }\href {https://doi.org/10.1038/nphys2672} {\bibfield  {journal} {\bibinfo
  {journal} {Nature Physics}\ }\textbf {\bibinfo {volume} {9}},\ \bibinfo
  {pages} {474} (\bibinfo {year} {2013})}\BibitemShut {NoStop}%
\bibitem [{\citenamefont {Shore}\ \emph {et~al.}(1989)\citenamefont {Shore},
  \citenamefont {Huang}, \citenamefont {Dorsey},\ and\ \citenamefont
  {Sethna}}]{shore_density_1989}%
  \BibitemOpen
  \bibfield  {author} {\bibinfo {author} {\bibfnamefont {J.~D.}\ \bibnamefont
  {Shore}}, \bibinfo {author} {\bibfnamefont {M.}~\bibnamefont {Huang}},
  \bibinfo {author} {\bibfnamefont {A.~T.}\ \bibnamefont {Dorsey}},\ and\
  \bibinfo {author} {\bibfnamefont {J.~P.}\ \bibnamefont {Sethna}},\ }\bibfield
   {title} {\bibinfo {title} {Density of states in a vortex core and the
  zero-bias tunneling peak},\ }\href
  {https://doi.org/10.1103/PhysRevLett.62.3089} {\bibfield  {journal} {\bibinfo
   {journal} {Physical Review Letters}\ }\textbf {\bibinfo {volume} {62}},\
  \bibinfo {pages} {3089} (\bibinfo {year} {1989})}\BibitemShut {NoStop}%
\bibitem [{\citenamefont {Xiang}\ \emph {et~al.}(2024)\citenamefont {Xiang},
  \citenamefont {Wang},\ and\ \citenamefont {Wang}}]{Xiang_SCPMA_2024}%
  \BibitemOpen
  \bibfield  {author} {\bibinfo {author} {\bibfnamefont {K.}~\bibnamefont
  {Xiang}}, \bibinfo {author} {\bibfnamefont {D.}~\bibnamefont {Wang}},\ and\
  \bibinfo {author} {\bibfnamefont {Q.-H.}\ \bibnamefont {Wang}},\ }\bibfield
  {title} {\bibinfo {title} {Quantized bound states around a vortex in
  anisotropic superconductors},\ }\href
  {https://doi.org/10.1007/s11433-023-2353-6} {\bibfield  {journal} {\bibinfo
  {journal} {SCIENCE CHINA Physics, Mechanics \& Astronomy}\ }\textbf {\bibinfo
  {volume} {67}},\ \bibinfo {pages} {267412} (\bibinfo {year}
  {2024})}\BibitemShut {NoStop}%
\bibitem [{\citenamefont {Nishimori}\ \emph {et~al.}(2004)\citenamefont
  {Nishimori}, \citenamefont {Uchiyama}, \citenamefont {Kaneko}, \citenamefont
  {Tokura}, \citenamefont {Takeya}, \citenamefont {Hirata},\ and\ \citenamefont
  {Nishida}}]{nishimori_first_2004}%
  \BibitemOpen
  \bibfield  {author} {\bibinfo {author} {\bibfnamefont {H.}~\bibnamefont
  {Nishimori}}, \bibinfo {author} {\bibfnamefont {K.}~\bibnamefont {Uchiyama}},
  \bibinfo {author} {\bibfnamefont {S.-i.}\ \bibnamefont {Kaneko}}, \bibinfo
  {author} {\bibfnamefont {A.}~\bibnamefont {Tokura}}, \bibinfo {author}
  {\bibfnamefont {H.}~\bibnamefont {Takeya}}, \bibinfo {author} {\bibfnamefont
  {K.}~\bibnamefont {Hirata}},\ and\ \bibinfo {author} {\bibfnamefont
  {N.}~\bibnamefont {Nishida}},\ }\bibfield  {title} {\bibinfo {title} {First
  {Observation} of the {Fourfold}-symmetric and {Quantum} {Regime} {Vortex}
  {Core} in {YNi2B2C} by {Scanning} {Tunneling} {Microscopy} and
  {Spectroscopy}},\ }\href {https://doi.org/10.1143/JPSJ.73.3247} {\bibfield
  {journal} {\bibinfo  {journal} {Journal of the Physical Society of Japan}\
  }\textbf {\bibinfo {volume} {73}},\ \bibinfo {pages} {3247} (\bibinfo {year}
  {2004})}\BibitemShut {NoStop}%
\bibitem [{\citenamefont {Chen}(2023)}]{chen_anisotropic_2023}%
  \BibitemOpen
  \bibfield  {author} {\bibinfo {author} {\bibfnamefont {H.-Y.}\ \bibnamefont
  {Chen}},\ }\bibfield  {title} {\bibinfo {title} {Anisotropic vortex core in
  the nematic state in electron-doped iron-pnictide superconductors},\ }\href
  {https://doi.org/10.1016/j.physc.2023.1354282} {\bibfield  {journal}
  {\bibinfo  {journal} {Physica C: Superconductivity and its Applications}\
  }\textbf {\bibinfo {volume} {610}},\ \bibinfo {pages} {1354282} (\bibinfo
  {year} {2023})}\BibitemShut {NoStop}%
\bibitem [{\citenamefont {Liu}\ \emph {et~al.}(2019)\citenamefont {Liu},
  \citenamefont {Tao}, \citenamefont {Ren}, \citenamefont {Chen}, \citenamefont
  {Yao}, \citenamefont {Wolf}, \citenamefont {Yan}, \citenamefont {Zhang},\
  and\ \citenamefont {Feng}}]{liu_evidence_2019}%
  \BibitemOpen
  \bibfield  {author} {\bibinfo {author} {\bibfnamefont {X.}~\bibnamefont
  {Liu}}, \bibinfo {author} {\bibfnamefont {R.}~\bibnamefont {Tao}}, \bibinfo
  {author} {\bibfnamefont {M.}~\bibnamefont {Ren}}, \bibinfo {author}
  {\bibfnamefont {W.}~\bibnamefont {Chen}}, \bibinfo {author} {\bibfnamefont
  {Q.}~\bibnamefont {Yao}}, \bibinfo {author} {\bibfnamefont {T.}~\bibnamefont
  {Wolf}}, \bibinfo {author} {\bibfnamefont {Y.}~\bibnamefont {Yan}}, \bibinfo
  {author} {\bibfnamefont {T.}~\bibnamefont {Zhang}},\ and\ \bibinfo {author}
  {\bibfnamefont {D.}~\bibnamefont {Feng}},\ }\bibfield  {title} {\bibinfo
  {title} {Evidence of nematic order and nodal superconducting gap along [110]
  direction in $\text{RbFe}_2\text{As}2$},\ }\href
  {https://www.nature.com/articles/s41467-019-08962-z} {\bibfield  {journal}
  {\bibinfo  {journal} {Nature Communications}\ }\textbf {\bibinfo {volume}
  {10}},\ \bibinfo {pages} {1039} (\bibinfo {year} {2019})}\BibitemShut
  {NoStop}%
\bibitem [{\citenamefont {Chen}\ \emph {et~al.}(2023)\citenamefont {Chen},
  \citenamefont {Chen}, \citenamefont {Wen}, \citenamefont {Hou}, \citenamefont
  {Yang},\ and\ \citenamefont {Wen}}]{chen_superconductivity_2023}%
  \BibitemOpen
  \bibfield  {author} {\bibinfo {author} {\bibfnamefont {K.}~\bibnamefont
  {Chen}}, \bibinfo {author} {\bibfnamefont {M.}~\bibnamefont {Chen}}, \bibinfo
  {author} {\bibfnamefont {C.}~\bibnamefont {Wen}}, \bibinfo {author}
  {\bibfnamefont {Z.}~\bibnamefont {Hou}}, \bibinfo {author} {\bibfnamefont
  {H.}~\bibnamefont {Yang}},\ and\ \bibinfo {author} {\bibfnamefont {H.-H.}\
  \bibnamefont {Wen}},\ }\bibfield  {title} {\bibinfo {title}
  {Superconductivity and vortex structure in
  $\text{Bi}_2\text{Te}_3$/$\text{FeTe}_{0.55}\text{Se}_{0.45}$
  heterostructures with varying $\text{Bi}_2\text{Te}_3$ thickness},\ }\href
  {https://doi.org/10.1103/PhysRevB.108.184512} {\bibfield  {journal} {\bibinfo
   {journal} {Physical Review B}\ }\textbf {\bibinfo {volume} {108}},\ \bibinfo
  {pages} {184512} (\bibinfo {year} {2023})}\BibitemShut {NoStop}%
\bibitem [{\citenamefont {Hayashi}\ \emph {et~al.}(1996)\citenamefont
  {Hayashi}, \citenamefont {Ichioka},\ and\ \citenamefont
  {Machida}}]{hayashi_star-shaped_1996}%
  \BibitemOpen
  \bibfield  {author} {\bibinfo {author} {\bibfnamefont {N.}~\bibnamefont
  {Hayashi}}, \bibinfo {author} {\bibfnamefont {M.}~\bibnamefont {Ichioka}},\
  and\ \bibinfo {author} {\bibfnamefont {K.}~\bibnamefont {Machida}},\
  }\bibfield  {title} {\bibinfo {title} {Star-{Shaped} {Local} {Density} of
  {States} around {Vortices} in a {Type}-{II} {Superconductor}},\ }\href
  {https://doi.org/10.1103/PhysRevLett.77.4074} {\bibfield  {journal} {\bibinfo
   {journal} {Phys. Rev. Lett.}\ }\textbf {\bibinfo {volume} {77}},\ \bibinfo
  {pages} {4074} (\bibinfo {year} {1996})}\BibitemShut {NoStop}%
\bibitem [{\citenamefont {Renner}\ \emph {et~al.}(1991)\citenamefont {Renner},
  \citenamefont {Kent}, \citenamefont {Niedermann}, \citenamefont {Fischer},\
  and\ \citenamefont {Levy}}]{renner_scanning_1991}%
  \BibitemOpen
  \bibfield  {author} {\bibinfo {author} {\bibfnamefont {C.}~\bibnamefont
  {Renner}}, \bibinfo {author} {\bibfnamefont {A.~D.}\ \bibnamefont {Kent}},
  \bibinfo {author} {\bibfnamefont {P.}~\bibnamefont {Niedermann}}, \bibinfo
  {author} {\bibfnamefont {O.}~\bibnamefont {Fischer}},\ and\ \bibinfo {author}
  {\bibfnamefont {F.}~\bibnamefont {Levy}},\ }\bibfield  {title} {\bibinfo
  {title} {Scanning tunneling spectroscopy of a vortex core from the clean to
  the dirty limit},\ }\href {https://doi.org/10.1103/PhysRevLett.67.1650}
  {\bibfield  {journal} {\bibinfo  {journal} {Physical Review Letters}\
  }\textbf {\bibinfo {volume} {67}},\ \bibinfo {pages} {1650} (\bibinfo {year}
  {1991})}\BibitemShut {NoStop}%
\bibitem [{\citenamefont {Miranovic}\ \emph {et~al.}(2004)\citenamefont
  {Miranovic}, \citenamefont {Ichioka},\ and\ \citenamefont
  {Machida}}]{miranovic_effects_2004}%
  \BibitemOpen
  \bibfield  {author} {\bibinfo {author} {\bibfnamefont {P.}~\bibnamefont
  {Miranovic}}, \bibinfo {author} {\bibfnamefont {M.}~\bibnamefont {Ichioka}},\
  and\ \bibinfo {author} {\bibfnamefont {K.}~\bibnamefont {Machida}},\
  }\bibfield  {title} {\bibinfo {title} {Effects of nonmagnetic scatterers on
  the local density of states around a vortex in $s$-wave superconductors},\
  }\href {https://doi.org/10.1103/PhysRevB.70.104510} {\bibfield  {journal}
  {\bibinfo  {journal} {Physical Review B}\ }\textbf {\bibinfo {volume} {70}},\
  \bibinfo {pages} {104510} (\bibinfo {year} {2004})}\BibitemShut {NoStop}%
\bibitem [{\citenamefont {Masaki}\ and\ \citenamefont
  {Kato}(2015)}]{masaki_impurity_2015}%
  \BibitemOpen
  \bibfield  {author} {\bibinfo {author} {\bibfnamefont {Y.}~\bibnamefont
  {Masaki}}\ and\ \bibinfo {author} {\bibfnamefont {Y.}~\bibnamefont {Kato}},\
  }\bibfield  {title} {\bibinfo {title} {Impurity {Effects} on {Bound} {States}
  in {Vortex} {Core} of {Topological} s-{Wave} {Superconductor}},\ }\href
  {https://doi.org/10.7566/JPSJ.84.094701} {\bibfield  {journal} {\bibinfo
  {journal} {Journal of the Physical Society of Japan}\ }\textbf {\bibinfo
  {volume} {84}},\ \bibinfo {pages} {094701} (\bibinfo {year}
  {2015})}\BibitemShut {NoStop}%
\bibitem [{\citenamefont {Tsutsumi}\ and\ \citenamefont
  {Kato}(2017)}]{tsutsumi_coherence_2017}%
  \BibitemOpen
  \bibfield  {author} {\bibinfo {author} {\bibfnamefont {Y.}~\bibnamefont
  {Tsutsumi}}\ and\ \bibinfo {author} {\bibfnamefont {Y.}~\bibnamefont
  {Kato}},\ }\bibfield  {title} {\bibinfo {title} {Coherence effects of
  caroli–de gennes–matricon modes in a nodal topological superconductor
  $\text{UPt}_3$},\ }\href {https://doi.org/10.1088/1742-6596/807/5/052005}
  {\bibfield  {journal} {\bibinfo  {journal} {Journal of Physics: Conference
  Series}\ }\textbf {\bibinfo {volume} {807}},\ \bibinfo {pages} {052005}
  (\bibinfo {year} {2017})}\BibitemShut {NoStop}%
\bibitem [{\citenamefont {Park}\ \emph {et~al.}(2021)\citenamefont {Park},
  \citenamefont {Barrena}, \citenamefont {Manas-Valero}, \citenamefont
  {Baldovi}, \citenamefont {Fente}, \citenamefont {Herrera}, \citenamefont
  {Mompean}, \citenamefont {Garcia-Hernandez}, \citenamefont {Rubio},
  \citenamefont {Coronado}, \citenamefont {Guillamon}, \citenamefont {Yeyati},\
  and\ \citenamefont {Suderow}}]{park_coherent_2021}%
  \BibitemOpen
  \bibfield  {author} {\bibinfo {author} {\bibfnamefont {S.}~\bibnamefont
  {Park}}, \bibinfo {author} {\bibfnamefont {V.}~\bibnamefont {Barrena}},
  \bibinfo {author} {\bibfnamefont {S.}~\bibnamefont {Manas-Valero}}, \bibinfo
  {author} {\bibfnamefont {J.~J.}\ \bibnamefont {Baldovi}}, \bibinfo {author}
  {\bibfnamefont {A.}~\bibnamefont {Fente}}, \bibinfo {author} {\bibfnamefont
  {E.}~\bibnamefont {Herrera}}, \bibinfo {author} {\bibfnamefont
  {F.}~\bibnamefont {Mompean}}, \bibinfo {author} {\bibfnamefont
  {M.}~\bibnamefont {Garcia-Hernandez}}, \bibinfo {author} {\bibfnamefont
  {A.}~\bibnamefont {Rubio}}, \bibinfo {author} {\bibfnamefont
  {E.}~\bibnamefont {Coronado}}, \bibinfo {author} {\bibfnamefont
  {I.}~\bibnamefont {Guillamon}}, \bibinfo {author} {\bibfnamefont {A.~L.}\
  \bibnamefont {Yeyati}},\ and\ \bibinfo {author} {\bibfnamefont
  {H.}~\bibnamefont {Suderow}},\ }\bibfield  {title} {\bibinfo {title}
  {Coherent coupling between vortex bound states and magnetic impurities in 2d
  layered superconductors},\ }\href
  {https://doi.org/10.1038/s41467-021-24531-9} {\bibfield  {journal} {\bibinfo
  {journal} {Nature Communications}\ }\textbf {\bibinfo {volume} {12}},\
  \bibinfo {pages} {4668} (\bibinfo {year} {2021})}\BibitemShut {NoStop}%
\bibitem [{\citenamefont {de~Mendonca}\ \emph {et~al.}(2023)\citenamefont
  {de~Mendonca}, \citenamefont {Manesco}, \citenamefont {Sandler},\ and\
  \citenamefont {Dias~da Silva}}]{de_mendonca_near_2023}%
  \BibitemOpen
  \bibfield  {author} {\bibinfo {author} {\bibfnamefont {B.~S.}\ \bibnamefont
  {de~Mendonca}}, \bibinfo {author} {\bibfnamefont {A.~L.~R.}\ \bibnamefont
  {Manesco}}, \bibinfo {author} {\bibfnamefont {N.}~\bibnamefont {Sandler}},\
  and\ \bibinfo {author} {\bibfnamefont {L.~G. G.~V.}\ \bibnamefont {Dias~da
  Silva}},\ }\bibfield  {title} {\bibinfo {title} {Near zero energy
  {Caroli}--de {Gennes}--{Matricon} vortex states in the presence of
  impurities},\ }\href {https://doi.org/10.1103/PhysRevB.107.184509} {\bibfield
   {journal} {\bibinfo  {journal} {Physical Review B}\ }\textbf {\bibinfo
  {volume} {107}},\ \bibinfo {pages} {184509} (\bibinfo {year}
  {2023})}\BibitemShut {NoStop}%
\bibitem [{\citenamefont {Wang}\ \emph {et~al.}(2018)\citenamefont {Wang},
  \citenamefont {Kong}, \citenamefont {Fan}, \citenamefont {Chen},
  \citenamefont {Zhu}, \citenamefont {Liu}, \citenamefont {Cao}, \citenamefont
  {Sun}, \citenamefont {Du}, \citenamefont {Schneeloch}, \citenamefont {Zhong},
  \citenamefont {Gu}, \citenamefont {Fu}, \citenamefont {Ding},\ and\
  \citenamefont {Gao}}]{exp-FeTeSe}%
  \BibitemOpen
  \bibfield  {author} {\bibinfo {author} {\bibfnamefont {D.}~\bibnamefont
  {Wang}}, \bibinfo {author} {\bibfnamefont {L.}~\bibnamefont {Kong}}, \bibinfo
  {author} {\bibfnamefont {P.}~\bibnamefont {Fan}}, \bibinfo {author}
  {\bibfnamefont {H.}~\bibnamefont {Chen}}, \bibinfo {author} {\bibfnamefont
  {S.}~\bibnamefont {Zhu}}, \bibinfo {author} {\bibfnamefont {W.}~\bibnamefont
  {Liu}}, \bibinfo {author} {\bibfnamefont {L.}~\bibnamefont {Cao}}, \bibinfo
  {author} {\bibfnamefont {Y.}~\bibnamefont {Sun}}, \bibinfo {author}
  {\bibfnamefont {S.}~\bibnamefont {Du}}, \bibinfo {author} {\bibfnamefont
  {J.}~\bibnamefont {Schneeloch}}, \bibinfo {author} {\bibfnamefont
  {R.}~\bibnamefont {Zhong}}, \bibinfo {author} {\bibfnamefont
  {G.}~\bibnamefont {Gu}}, \bibinfo {author} {\bibfnamefont {L.}~\bibnamefont
  {Fu}}, \bibinfo {author} {\bibfnamefont {H.}~\bibnamefont {Ding}},\ and\
  \bibinfo {author} {\bibfnamefont {H.-J.}\ \bibnamefont {Gao}},\ }\bibfield
  {title} {\bibinfo {title} {Evidence for majorana bound states in an
  iron-based superconductor},\ }\href {https://doi.org/10.1126/science.aao1797}
  {\bibfield  {journal} {\bibinfo  {journal} {Science}\ }\textbf {\bibinfo
  {volume} {362}},\ \bibinfo {pages} {333} (\bibinfo {year}
  {2018})}\BibitemShut {NoStop}%
\bibitem [{\citenamefont {Liu}\ \emph {et~al.}(2018)\citenamefont {Liu},
  \citenamefont {Chen}, \citenamefont {Zhang}, \citenamefont {Peng},
  \citenamefont {Yan}, \citenamefont {Wen}, \citenamefont {Lou}, \citenamefont
  {Huang}, \citenamefont {Tian}, \citenamefont {Dong}, \citenamefont {Wang},
  \citenamefont {Bao}, \citenamefont {Wang}, \citenamefont {Yin}, \citenamefont
  {Zhao},\ and\ \citenamefont {Feng}}]{exp-LiFeOHFeSe}%
  \BibitemOpen
  \bibfield  {author} {\bibinfo {author} {\bibfnamefont {Q.}~\bibnamefont
  {Liu}}, \bibinfo {author} {\bibfnamefont {C.}~\bibnamefont {Chen}}, \bibinfo
  {author} {\bibfnamefont {T.}~\bibnamefont {Zhang}}, \bibinfo {author}
  {\bibfnamefont {R.}~\bibnamefont {Peng}}, \bibinfo {author} {\bibfnamefont
  {Y.-J.}\ \bibnamefont {Yan}}, \bibinfo {author} {\bibfnamefont {C.-H.-P.}\
  \bibnamefont {Wen}}, \bibinfo {author} {\bibfnamefont {X.}~\bibnamefont
  {Lou}}, \bibinfo {author} {\bibfnamefont {Y.-L.}\ \bibnamefont {Huang}},
  \bibinfo {author} {\bibfnamefont {J.-P.}\ \bibnamefont {Tian}}, \bibinfo
  {author} {\bibfnamefont {X.-L.}\ \bibnamefont {Dong}}, \bibinfo {author}
  {\bibfnamefont {G.-W.}\ \bibnamefont {Wang}}, \bibinfo {author}
  {\bibfnamefont {W.-C.}\ \bibnamefont {Bao}}, \bibinfo {author} {\bibfnamefont
  {Q.-H.}\ \bibnamefont {Wang}}, \bibinfo {author} {\bibfnamefont {Z.-P.}\
  \bibnamefont {Yin}}, \bibinfo {author} {\bibfnamefont {Z.-X.}\ \bibnamefont
  {Zhao}},\ and\ \bibinfo {author} {\bibfnamefont {D.-L.}\ \bibnamefont
  {Feng}},\ }\bibfield  {title} {\bibinfo {title} {Robust and clean majorana
  zero mode in the vortex core of high-temperature superconductor
  $\mathbf{(}{\mathrm{li}}_{0.84}{\mathrm{fe}}_{0.16}\mathbf{)}\mathrm{OHFeSe}$},\
  }\href {https://doi.org/10.1103/PhysRevX.8.041056} {\bibfield  {journal}
  {\bibinfo  {journal} {Phys. Rev. X}\ }\textbf {\bibinfo {volume} {8}},\
  \bibinfo {pages} {041056} (\bibinfo {year} {2018})}\BibitemShut {NoStop}%
\bibitem [{\citenamefont {Kong}\ \emph {et~al.}(2021)\citenamefont {Kong},
  \citenamefont {Cao}, \citenamefont {Zhu}, \citenamefont {Papaj},
  \citenamefont {Dai}, \citenamefont {Li}, \citenamefont {Fan}, \citenamefont
  {Liu}, \citenamefont {Yang}, \citenamefont {Wang}, \citenamefont {Du},
  \citenamefont {Jin}, \citenamefont {Fu}, \citenamefont {Gao},\ and\
  \citenamefont {Ding}}]{exp-LiFeAs}%
  \BibitemOpen
  \bibfield  {author} {\bibinfo {author} {\bibfnamefont {L.}~\bibnamefont
  {Kong}}, \bibinfo {author} {\bibfnamefont {L.}~\bibnamefont {Cao}}, \bibinfo
  {author} {\bibfnamefont {S.}~\bibnamefont {Zhu}}, \bibinfo {author}
  {\bibfnamefont {M.}~\bibnamefont {Papaj}}, \bibinfo {author} {\bibfnamefont
  {G.}~\bibnamefont {Dai}}, \bibinfo {author} {\bibfnamefont {G.}~\bibnamefont
  {Li}}, \bibinfo {author} {\bibfnamefont {P.}~\bibnamefont {Fan}}, \bibinfo
  {author} {\bibfnamefont {W.}~\bibnamefont {Liu}}, \bibinfo {author}
  {\bibfnamefont {F.}~\bibnamefont {Yang}}, \bibinfo {author} {\bibfnamefont
  {X.}~\bibnamefont {Wang}}, \bibinfo {author} {\bibfnamefont {S.}~\bibnamefont
  {Du}}, \bibinfo {author} {\bibfnamefont {C.}~\bibnamefont {Jin}}, \bibinfo
  {author} {\bibfnamefont {L.}~\bibnamefont {Fu}}, \bibinfo {author}
  {\bibfnamefont {H.-J.}\ \bibnamefont {Gao}},\ and\ \bibinfo {author}
  {\bibfnamefont {H.}~\bibnamefont {Ding}},\ }\bibfield  {title} {\bibinfo
  {title} {Majorana zero modes in impurity-assisted vortex of {LiFeAs}
  superconductor},\ }\href {https://doi.org/10.1038/s41467-021-24372-6}
  {\bibfield  {journal} {\bibinfo  {journal} {Nature Communications}\ }\textbf
  {\bibinfo {volume} {12}},\ \bibinfo {pages} {4146} (\bibinfo {year}
  {2021})}\BibitemShut {NoStop}%
\bibitem [{\citenamefont {Nayak}\ \emph {et~al.}(2008)\citenamefont {Nayak},
  \citenamefont {Simon}, \citenamefont {Stern}, \citenamefont {Freedman},\ and\
  \citenamefont {Das~Sarma}}]{Nayak2008}%
  \BibitemOpen
  \bibfield  {author} {\bibinfo {author} {\bibfnamefont {C.}~\bibnamefont
  {Nayak}}, \bibinfo {author} {\bibfnamefont {S.~H.}\ \bibnamefont {Simon}},
  \bibinfo {author} {\bibfnamefont {A.}~\bibnamefont {Stern}}, \bibinfo
  {author} {\bibfnamefont {M.}~\bibnamefont {Freedman}},\ and\ \bibinfo
  {author} {\bibfnamefont {S.}~\bibnamefont {Das~Sarma}},\ }\bibfield  {title}
  {\bibinfo {title} {Non-abelian anyons and topological quantum computation},\
  }\href {https://doi.org/10.1103/RevModPhys.80.1083} {\bibfield  {journal}
  {\bibinfo  {journal} {Rev. Mod. Phys.}\ }\textbf {\bibinfo {volume} {80}},\
  \bibinfo {pages} {1083} (\bibinfo {year} {2008})}\BibitemShut {NoStop}%
\bibitem [{\citenamefont {Mel'nikov}\ \emph {et~al.}(2008)\citenamefont
  {Mel'nikov}, \citenamefont {Ryzhov},\ and\ \citenamefont
  {Silaev}}]{melnikov_electronic_2008}%
  \BibitemOpen
  \bibfield  {author} {\bibinfo {author} {\bibfnamefont {A.~S.}\ \bibnamefont
  {Mel'nikov}}, \bibinfo {author} {\bibfnamefont {D.~A.}\ \bibnamefont
  {Ryzhov}},\ and\ \bibinfo {author} {\bibfnamefont {M.~A.}\ \bibnamefont
  {Silaev}},\ }\bibfield  {title} {\bibinfo {title} {Electronic structure and
  heat transport of multivortex configurations in mesoscopic superconductors},\
  }\href {https://doi.org/10.1103/PhysRevB.78.064513} {\bibfield  {journal}
  {\bibinfo  {journal} {Phys. Rev. B}\ }\textbf {\bibinfo {volume} {78}},\
  \bibinfo {pages} {064513} (\bibinfo {year} {2008})}\BibitemShut {NoStop}%
\bibitem [{\citenamefont {Mel'nikov}\ \emph {et~al.}(2009)\citenamefont
  {Mel'nikov}, \citenamefont {Ryzhov},\ and\ \citenamefont
  {Silaev}}]{melnikov_local_2009}%
  \BibitemOpen
  \bibfield  {author} {\bibinfo {author} {\bibfnamefont {A.~S.}\ \bibnamefont
  {Mel'nikov}}, \bibinfo {author} {\bibfnamefont {D.~A.}\ \bibnamefont
  {Ryzhov}},\ and\ \bibinfo {author} {\bibfnamefont {M.~A.}\ \bibnamefont
  {Silaev}},\ }\bibfield  {title} {\bibinfo {title} {Local density of states
  around single vortices and vortex pairs: {Effect} of boundaries and
  hybridization of vortex core states},\ }\href
  {https://doi.org/10.1103/PhysRevB.79.134521} {\bibfield  {journal} {\bibinfo
  {journal} {Phys. Rev. B}\ }\textbf {\bibinfo {volume} {79}},\ \bibinfo
  {pages} {134521} (\bibinfo {year} {2009})}\BibitemShut {NoStop}%
\bibitem [{\citenamefont {Franz}\ and\ \citenamefont
  {Tesanovic}(2000)}]{franz_quasiparticles_2000}%
  \BibitemOpen
  \bibfield  {author} {\bibinfo {author} {\bibfnamefont {M.}~\bibnamefont
  {Franz}}\ and\ \bibinfo {author} {\bibfnamefont {Z.}~\bibnamefont
  {Tesanovic}},\ }\bibfield  {title} {\bibinfo {title} {Quasiparticles in the
  {Vortex} {Lattice} of {Unconventional} {Superconductors}: {Bloch} {Waves} or
  {Landau} {Levels}?},\ }\href {https://doi.org/10.1103/PhysRevLett.84.554}
  {\bibfield  {journal} {\bibinfo  {journal} {Phys. Rev. Lett.}\ }\textbf
  {\bibinfo {volume} {84}},\ \bibinfo {pages} {554} (\bibinfo {year}
  {2000})}\BibitemShut {NoStop}%
\bibitem [{\citenamefont {Schopohl}\ and\ \citenamefont
  {Maki}(1995)}]{Schopohl1995}%
  \BibitemOpen
  \bibfield  {author} {\bibinfo {author} {\bibfnamefont {N.}~\bibnamefont
  {Schopohl}}\ and\ \bibinfo {author} {\bibfnamefont {K.}~\bibnamefont
  {Maki}},\ }\bibfield  {title} {\bibinfo {title} {Quasiparticle spectrum
  around a vortex line in a d-wave superconductor},\ }\href
  {https://doi.org/10.1103/PhysRevB.52.490} {\bibfield  {journal} {\bibinfo
  {journal} {Phys. Rev. B}\ }\textbf {\bibinfo {volume} {52}},\ \bibinfo
  {pages} {490} (\bibinfo {year} {1995})}\BibitemShut {NoStop}%
\bibitem [{\citenamefont {Silaev}\ and\ \citenamefont
  {Silaeva}(2013)}]{silaev_self-consistent_2013}%
  \BibitemOpen
  \bibfield  {author} {\bibinfo {author} {\bibfnamefont {M.~A.}\ \bibnamefont
  {Silaev}}\ and\ \bibinfo {author} {\bibfnamefont {V.~A.}\ \bibnamefont
  {Silaeva}},\ }\bibfield  {title} {\bibinfo {title} {Self-consistent
  electronic structure of multiquantum vortices in superconductors at
  {T}$\ll${Tc}},\ }\href {https://doi.org/10.1088/0953-8984/25/22/225702}
  {\bibfield  {journal} {\bibinfo  {journal} {J. Phys. Condens. Matter}\
  }\textbf {\bibinfo {volume} {25}},\ \bibinfo {pages} {225702} (\bibinfo
  {year} {2013})}\BibitemShut {NoStop}%
\end{thebibliography}%
\end{document}